\documentclass[a4paper,11pt]{article}
\usepackage{pos}

\usepackage{cleveref}

\newcommand{\ket}[1]{{|#1\rangle}}
\newcommand{\cHT}{{{\cal H}_{\mathrm{Trad}}}}
\newcommand{\cH}{{\cal H}}
\newcommand{\bS}{{\mathbf{S}}}
\newcommand{\mdc}{\{\lambda_\ell\},\{\lambda_s\},\{\alpha_s\}}
\newcommand{\chE}{\hat{{\cal E}}}

\renewcommand{\logo}{\relax}

\title{Qubit Regularization of Quantum Field Theories}

\author*[a]{Shailesh Chandrasekharan}

\affiliation[a]{Department of Physics, Duke University,\\
Box 90305, Duke University, Durham, NC 27708, USA}

\emailAdd{sch27@duke.edu}

\abstract{To study quantum field theories on a quantum computer, we must begin with Hamiltonians defined on a finite-dimensional Hilbert space and then take appropriate limits. This approach can be seen as a new type of regularization for quantum field theories, which we refer to as qubit regularization. A related finite-dimensional regularization, known as the D-theory approach, was proposed long ago as a general framework for all quantum field theories. In this framework, the dimensionality of the local Hilbert space at each spatial point can increase as needed through an additional flavor index. To reproduce asymptotically free QFTs, most studies assume that qubit-regularized theories require extending the local Hilbert space to infinity. However, contrary to this common belief, recent discoveries in (1+1) dimensions have revealed two examples where asymptotic freedom appears to emerge within a strictly finite-dimensional local Hilbert space through a novel renormalization group (RG) flow. These findings motivate further investigation into whether asymptotically free gauge theories could also emerge within a strictly finite-dimensional local Hilbert space. To support these explorations, we propose an orthonormal basis called the monomer-dimer-tensor-network (MDTN) basis and use it to construct new types of qubit-regularized lattice gauge theories.
}

\FullConference{The 41st International Symposium on Lattice Field Theory (LATTICE2024)\\
 28 July - 3 August 2024\\
Liverpool, UK\\}

\begin{document}
\maketitle

\section{Introduction}
\label{sec1}

The possibility of using quantum computers to solve quantum field theories (QFTs) provides an opportunity to investigate how these theories, traditionally constructed on infinite-dimensional Hilbert spaces, can emerge as limits of finite quantum mechanical systems \cite{Qsim2023,Qcmp2023}. This finite-dimensional, matrix model approach to QFTs holds the potential to reveal deeper insights into the underlying physics, going beyond its applications in quantum computation.

We refer to this finite-dimensional regularization of QFTs as qubit regularization. While traditional lattice regularization provides a starting point, the infinite-dimensional local Hilbert space of bosonic quantum fields requires further regularization. The D-theory proposed such a finite dimensional formulation for many QFTs including gauge theories nearly two decades ago \cite{DTh2004}. In that approach, an extra dimension (or equivalently a flavor index) was introduced at every spatial lattice point, allowing for a systematic increase in the local Hilbert space when necessary. This philosophy has also inspired many recent studies, which often assume that the local Hilbert space will ultimately need to be extended to infinity to formulate asymptotically free QFTs, such as Yang-Mills theories and QCD.

\begin{figure}[b]
\centering \includegraphics[width=0.8\textwidth]{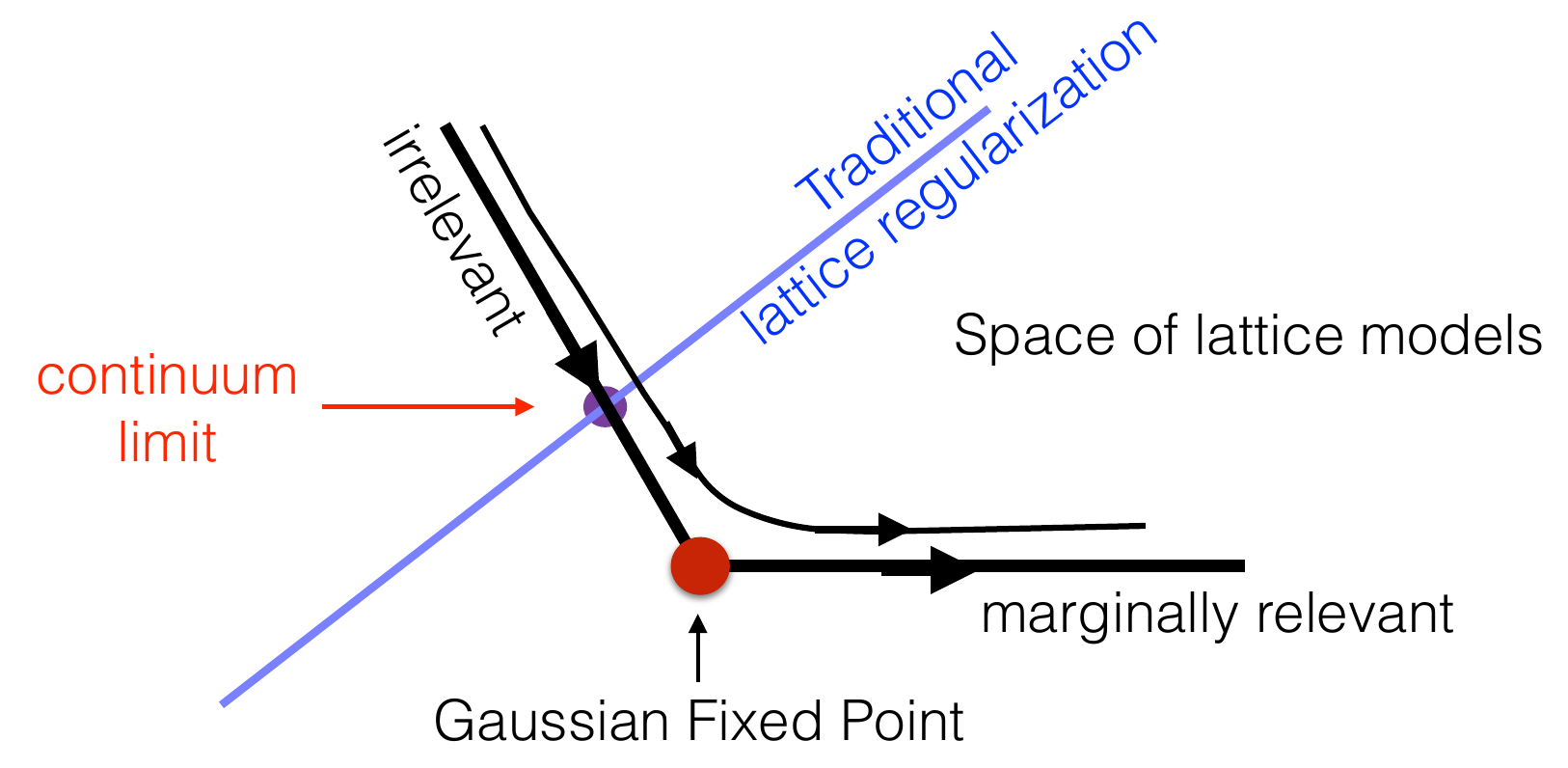}
\caption{Traditional RG flow in the space of lattice models for an asymptotically free QFT with a Gaussian UV fixed point (FP). It is straightforward to construct lattice models that flow to the Gaussian FP, if the local Hilbert space is allowed to be infinite-dimensional. However, this becomes challenging with qubit regularization.}
  \label{fig1}
\end{figure}

Typical asymptotically free QFTs can be viewed as massive theories emerging from a free (Gaussian) UV fixed point via a marginally relevant coupling. A schematic of the traditional RG flow in such theories is shown in \cref{fig1}. While constructing lattice theories with infinite-dimensional local Hilbert spaces that flow to the desired UV Gaussian fixed point is straightforward, achieving this within a strictly finite Hilbert space is more challenging, as it minimally requires fine-tuning to reach the critical surface. This fine-tuning approach, largely unexplored, represents a new research direction motivated by quantum computation.

One of the goals of this talk is to argue for the possibility of a novel non-perturbative RG flow through which qubit-regularized quantum field theories may recover asymptotically free QFTs. Specifically, we will discuss two examples of qubit regularization that provide concrete evidence for these new RG flows. First, we will summarize a recent study \cite{AFPRL2021} demonstrating how the asymptotically free fixed point of the two-dimensional O(3) model can emerge using a local four-dimensional Hilbert space in the Hamiltonian formulation. Next, we will review another study \cite{AFPRL2024} showing how the massive QFT arising at the BKT transition can emerge from a simple four-dimensional Hilbert space in the Lagrangian formulation. Both examples highlight the possibility that asymptotic freedom may emerge through new types of RG flows in qubit-regularized theories.

\begin{figure}[t]
\centering \includegraphics[width=0.8\textwidth]{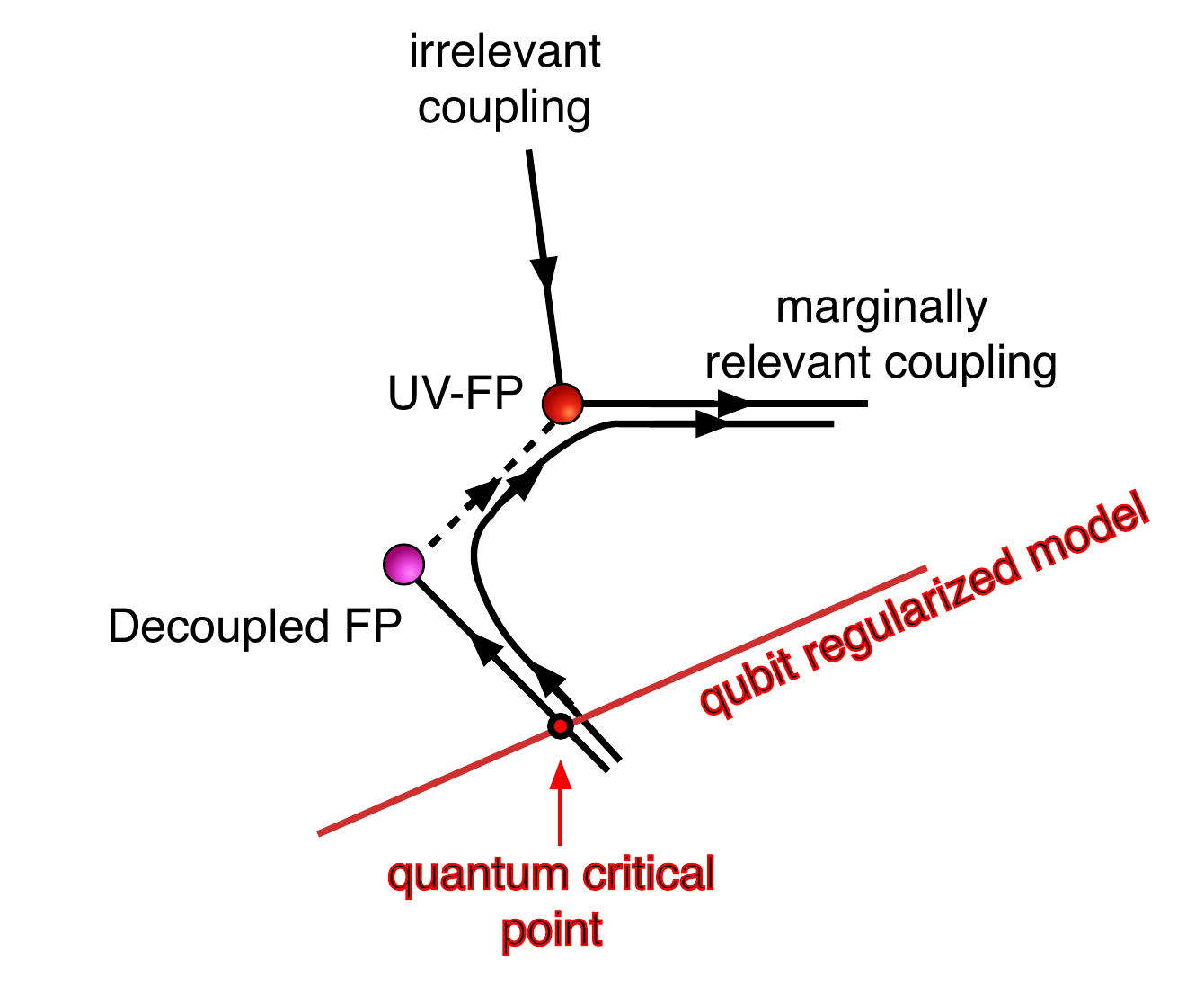}
\caption{This figure illustrates an alternative RG flow, discovered at the critical points of two qubit-regularized field theories, demonstrating how asymptotically free QFTs with a UV fixed point (FP) can emerge as a crossover critical phenomenon, while the RG flow at the critical point itself leads to a completely different decoupled FP.}
  \label{fig2}
\end{figure}

To understand this new RG flow in the two examples, consider a qubit regularized quantum mechanical Hamiltonian acting on a finite local lattice Hilbert space. Such a system typically depends on a lattice size \( L \) and a set of couplings \( g \), with the desired QFT emerging in the limits \( g \rightarrow g_c \) and \( L \rightarrow \infty \). However, the physics of the QFT can be obscured by the implementation of these limiting procedures. While the traditional RG flow diagram in \cref{fig1} suggests that setting \( g = g_c \) recovers the Gaussian ultraviolet (UV) fixed point, the new examples indicate that, instead, the lattice theory flows to a completely different fixed point — one we refer to as a decoupled fixed point, based on the observed physics in these cases. Nevertheless, as \( g \) approaches \( g_c \), all the universal physics of the desired UV fixed point can still be recovered as a crossover phenomenon. Specifically, for small \( L \), the theory is dominated by the decoupled fixed point; for intermediate \( L \), the physics of the desired UV fixed point becomes visible; and for very large \( L \), the theory exhibits the universal behavior of the massive QFT. This distinct and novel RG flow is illustrated in \cref{fig2}.

\section{Qubit Regularization of the Asymptotically Free \texorpdfstring{$O(3)$}{O(3)} QFT in Two Dimensions}
\label{sec2}

In this section, we review the main results of Ref.~\cite{AFPRL2021}, which demonstrate how the qubit regularization of the asymptotically free \( O(3) \) quantum field theory in two dimensions (2D \( O(3) \) QFT) can be achieved using only two qubits per lattice site. We also argue that the RG flow, which recovers asymptotic freedom in the UV, is given by \cref{fig2}.

The traditional formulations of the 2D \( O(3) \) QFT begins with an infinite-dimensional local Hilbert space \( \cHT \) at each lattice site, representing a quantum particle constrained to move on the surface of a unit sphere in three dimensions. The position of the particle is described by the unit vector \( \vec{\phi} \), and the corresponding quantum eigenstates \( \ket{\vec{\phi}} \) form a complete basis for the Hilbert space. 

One of the many ways to quantitatively understand the asymptotic freedom of the 2D \( O(3) \) QFT is by defining a finite-volume correlation length \( \xi(L) \) and computing it as a function of the box size \( L \). Using \( \xi(L) \), we can then compute the step-scaling function (SSF) \( f(x) \), where \( x = \xi(L)/L \) and \( f(x) = \xi(2L)/\xi(L) \). The function \( f(x) \) is a nonlinear function and is well defined for all values of \( x \), with the limiting values \( f(x \to \infty) = 2 \) (UV regime) and \( f(x \to 0) = 1 \) (IR regime). 

\begin{figure}[t]
\centering \includegraphics[width=0.8\textwidth]{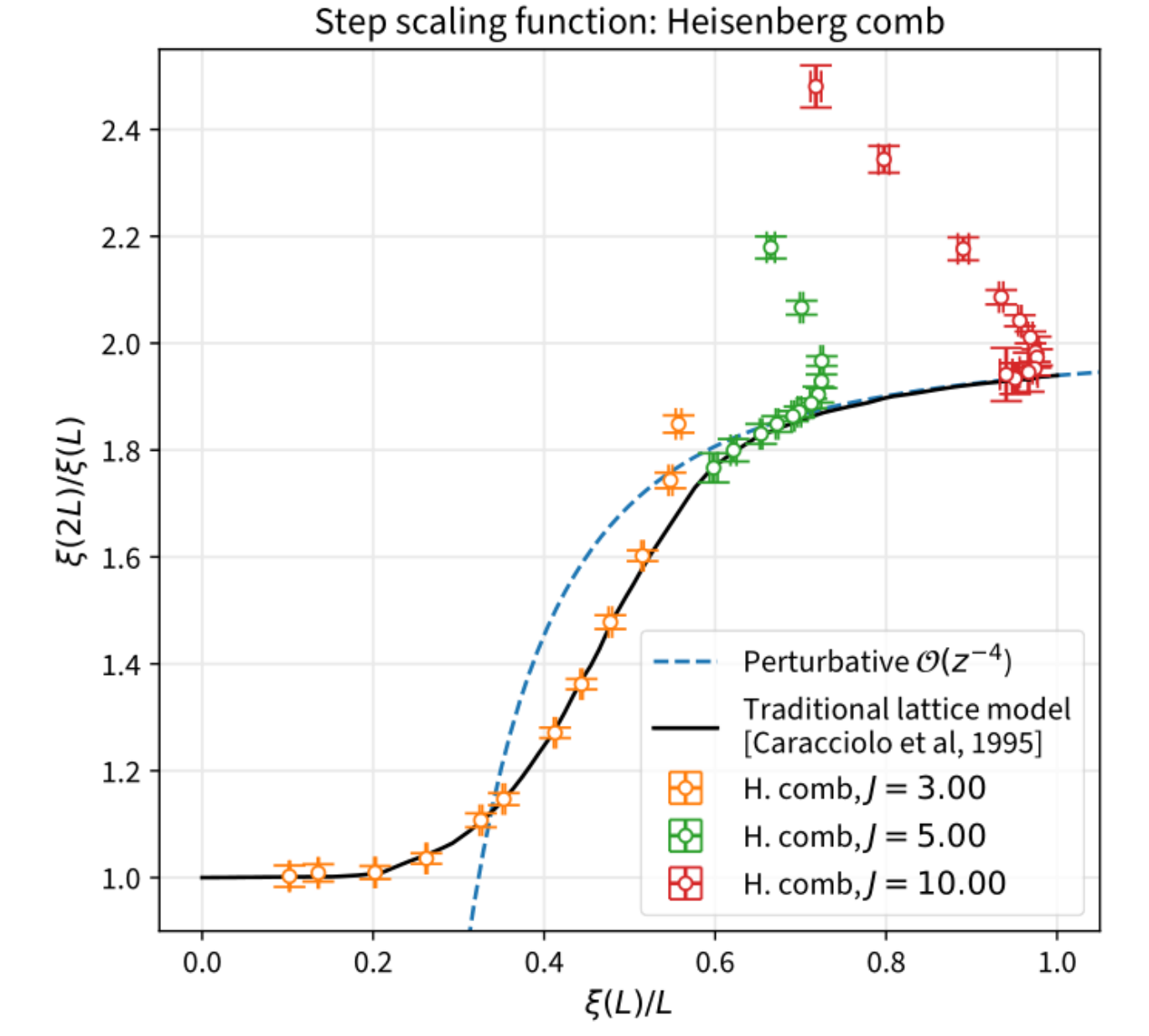}
\caption{The SSF of the 2D \( O(3) \) QFT. The solid black line represents the results obtained in Ref.~\cite{SSFO31995} using the traditional lattice formulation, while the data points correspond to the Heisenberg comb Hamiltonian described in \cref{hcomb}. For each fixed value of $J$, the data for $L > L_{\rm min}$ begins to align with the universal SSF, and $L_{\rm min}$ increases with $J$, allowing us to access more of the SSF in the UV regime. This figure was originally published in \cite{AFPRL2021}.}
  \label{fig3}
\end{figure}

For one of the many possible definitions of \( \xi(L) \) using the correlations of $\vec{\phi}$, the SSF was computed non-perturbatively using the traditional Lagrangian lattice formulation of the 2D \( O(3) \) QFT in Ref. \cite{SSFO31995} and is shown as the black solid line in \cref{fig3}. The perturbative result, starting from the Gaussian UV fixed point, is also shown as a dashed line.

A key challenge in qubit regularization is to reproduce this SSF of the traditional formulation using a lattice model with a finite-dimensional Hilbert space. Symmetries can provide valuable guidance in this process. Since the 2D \( O(3) \) QFT exhibits \( SO(3) \) symmetry, it is natural to preserve this symmetry under qubit regularization. A natural approach is to decompose the traditional local Hilbert space at each lattice site as a direct sum over the irreducible representations (irreps) of \( SO(3) \):
\begin{align}
\cHT = \bigoplus_{\ell=0,1,2,\ldots} \cH_\ell,
\label{tradH}
\end{align}
where \( \cH_\ell \) denotes the irreducible representation of \( SO(3) \) with angular momentum \( \ell \). Each Hilbert space \( \cH_\ell \) has dimension \( 2\ell+1 \), with basis states labeled by the standard orbital angular momentum states \( \ket{\ell, m} \), where \( -\ell \leq m \leq \ell \).

A simple type of qubit regularization begins with a lattice model defined on a truncated Hilbert space, constructed using only a few allowed values of $\ell$ in the sum \cref{tradH}. For example, in Ref.~\cite{AFPRL2024}, the qubit-regularized Hilbert space chosen at each lattice site was  
\begin{align}
\cH_Q = \cH_{\ell = 0} \oplus \cH_{\ell=1}.
\end{align}  
This four-dimensional space was implemented using two qubits per lattice site. Denoting the corresponding spin-\(\frac{1}{2}\) operators as  
\( \bS_{x,1} \) and \( \bS_{x,2} \), the lattice Hamiltonian of the qubit-regularized model in one spatial dimension is given by  
\begin{align}
H \ =\  \sum_{x=0}^{L-1} \ J \ \bS_{x,1} \cdot \bS_{x+1,1} \ +\  \bS_{x,1} \cdot \bS_{x,2}.
\label{hcomb}
\end{align}
This model is referred to as the {\em Heisenberg comb}. Using antiferromagnetic spin-spin correlation functions, $\xi(L)$ can be calculated as a function of $L$ and which can then be used to compute the SSF. As we will argue below, this SSF is exactly the same as the one obtained in the traditional model in the \( J \to \infty \) limit, in the appropriate regime of lattice sizes.

It is well known that, in order to reproduce a continuum QFT, the lattice model needs to be tuned to a critical point. In the Heisenberg comb, \( J=\infty \) is one such critical point.  When \( J=\infty \), the spins \( \bS_{x,2} \) decouple from the spins \( \bS_{x,1} \), which form a spin-\(\frac{1}{2}\) chain that is known to be critical. In the infrared (IR), this chain flows to the \( k=1 \) Wess-Zumino-Witten (WZW) conformal field theory. 

Thus, at the \( J=\infty \) critical point, the Heisenberg comb flows to the \( k=1 \) WZW fixed point along with an infinite number of decoupled spins \( \bS_{x,2} \). This is the decoupled fixed point shown in \cref{fig2}. As expected, for very large but finite values of \( J \) (i.e., \( J \neq \infty \)), the model no longer flows to the decoupled fixed point and instead becomes massive. Interestingly, however, the RG flow takes it arbitrarily close to the Gaussian UV fixed point.  

This flow can be analyzed by computing \( x = \xi(L)/L \) and \( y = \xi(2L)/\xi(L) \), and plotting these points on an \( (x,y) \) graph for various values of \( L \) at a fixed \( J \). This allows us to observe the pattern of the RG flow from the UV to the infrared (IR) as \( L \) increases.  

Using large-scale quantum Monte Carlo algorithms, this behavior can be studied in detail and is shown in \cref{fig3} for \( J=3,5,10 \), and \( 12 \leq L \leq 512 \). For \( L < L_{\rm min} \), the plotted Monte Carlo data do not exhibit any recognizable scaling pattern. However, when \( L > L_{\rm min} \), they begin to align with the expected step-scaling function (SSF) for the 2D \( O(3) \) QFT, represented by the black solid line.  

Note also that \( L_{\rm min}(J) \) is a function of \( J \) and increases with increasing \( J \). Additionally, the ratio \( \xi(L_{\rm min})/L_{\rm min} \) also increases, suggesting that as \( J \) grows, the qubit model captures more of the ultraviolet (UV) physics of the 2D \( O(3) \) QFT accurately.  

This indicates that the Gaussian UV fixed point of the 2D \( O(3) \) QFT emerges as a crossover phenomenon in the Heisenberg comb and can be fully recovered in the \( J \to \infty \) limit by also focusing on lattice sizes \( L > L_{\rm min}(J) \). The actual RG flow of the Heisenberg comb appears to be best described by the novel RG flow shown in \cref{fig2}.

\section{Qubit regularization of the massive QFT at the BKT critical point}
\label{sec3}

In this section, we review the main results of Ref.~\cite{AFPRL2024}, which demonstrate how the massive QFT at the Berezinskii-Kosterlitz-Thouless (BKT) critical point, defined through the traditional XY model, can be reproduced using a qubit-regularized model with a four-dimensional local Hilbert space. We also find once again that the RG flow, which recovers the continuum physics of the QFT, is given by \cref{fig2}.

The traditional XY model is defined using an infinite-dimensional local Hilbert space \( \cHT \) at each lattice site, corresponding to a quantum particle moving on a circle of unit radius. The position of the particle is described by the angle \( 0 \leq \theta < 2\pi \). The action of the model on a two-dimensional square lattice, representing Euclidean space-time, is given by  
\begin{align}
S = \beta \sum_{\langle xy\rangle} \cos(\theta_x - \theta_y),
\end{align}  
where \( \beta_c \approx 1.1199(1) \) is the Berezinskii-Kosterlitz-Thouless (BKT) critical point \cite{BXY2005}. For \( \beta < \beta_c \), the lattice model is in a massive phase, where the correlation length grows exponentially as \( \beta \to \beta_c \). At \( \beta_c \), the infrared (IR) quantum field theory (QFT) consists of free bosons. 

\begin{figure}[t]
\centering \includegraphics[width=\textwidth]{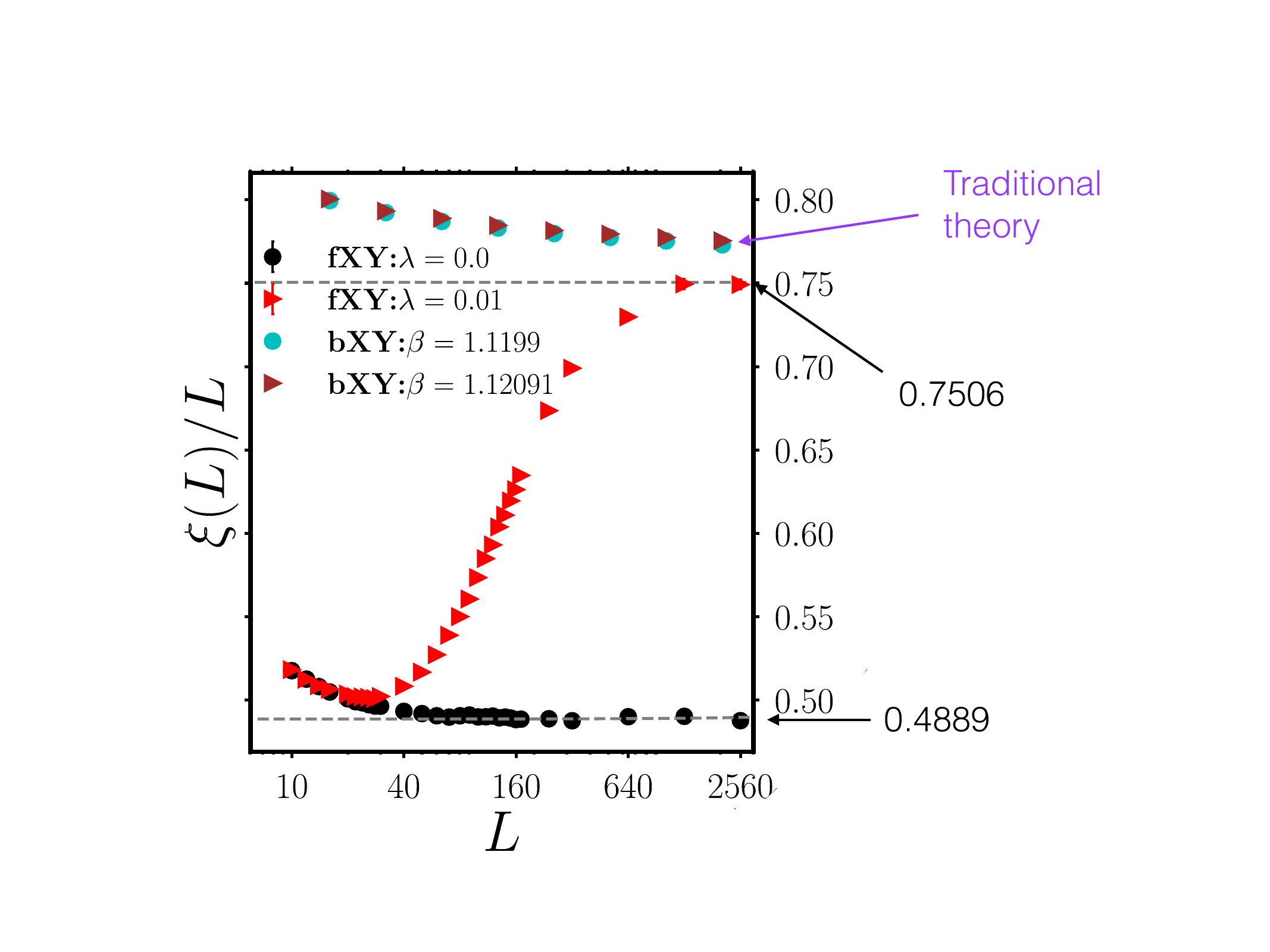}
\caption{The plot of $\xi(L)/L$ as a function of $L$ in the traditional XY model (referred to as bXY in the figure) as compared to the qubit regularized model (referred to as the fXY in the figure). Note that $\xi(L)/L$ does not reach the value of $0.7506...$ expected in the traditional XY model at the BKT transition for $\beta_c\approx 1.1199(1)$ even at $L\approx 2500$, while the qubit regularized model recovers this quite accurately when $\lambda=0.01$ at those lattice sizes. In contrast, when $\lambda=0$ the qubit regularized theory is very different since $\xi(L)/L \approx 0.4889$. \label{fig6}}
\end{figure}

These features motivate the characterization of the massive continuum QFT that emerges from the lattice XY model in the limit \( \beta \to \beta_c \) as an asymptotically free QFT. We will refer to this theory as the 2D \( O(2) \) QFT. As discussed in \cref{sec2}, we can once again define an SSF for this 2D \( O(2) \) QFT by defining \( \xi(L) \) through correlations of \( e^{i\theta} \) and \( e^{-i\theta} \). This SSF was computed in Ref.~\cite{AFPRL2024} and is represented by the solid lines in the four graphs shown in \cref{fig5}. At the BKT critical point, we expect \( \xi(L)/L =  0.7506912\ldots \) for large values of \( L \) \cite{BXY2005}. However, the traditional model does not reach this value even when \( L \approx 2500 \), as seen in \cref{fig6}. This discrepancy is usually attributed to slowly varying logarithmic finite volume corrections.

\begin{figure}[t]
\centering \includegraphics[width=\textwidth]{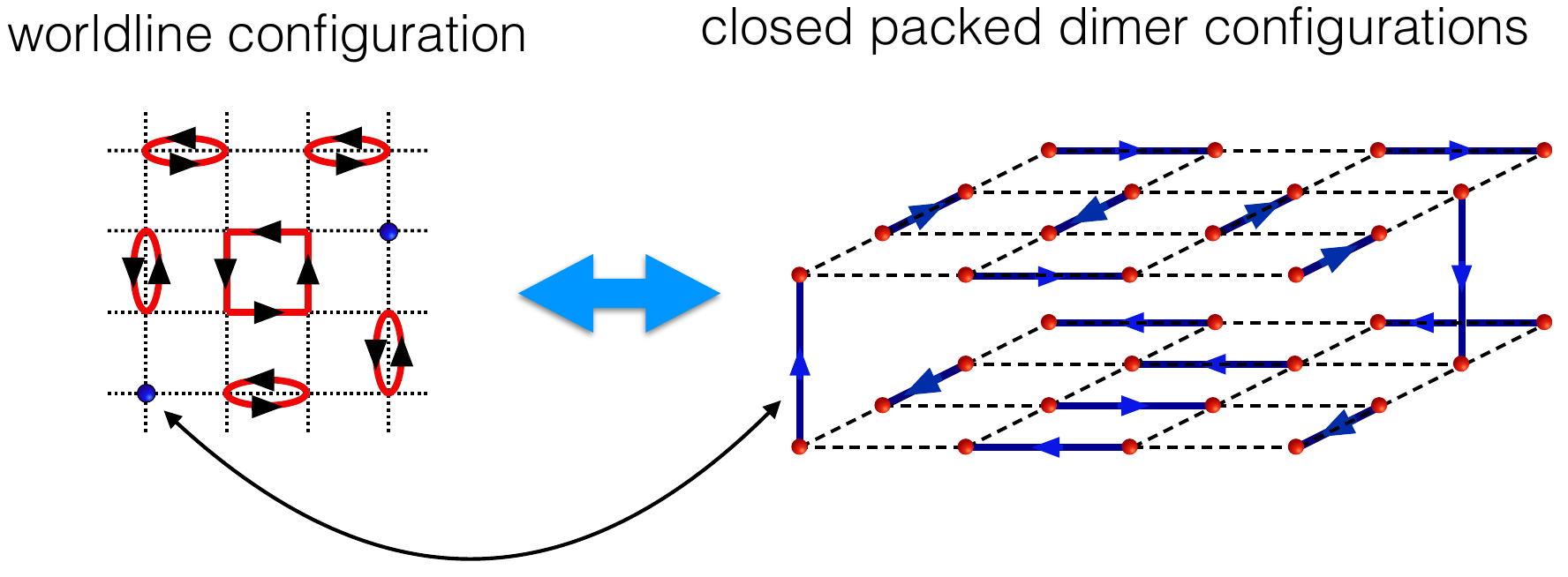}
\caption{Illustration of configurations \( C \) that define the qubit-regularized model in \cref{dimer}. Each worldline configuration (left) can be uniquely mapped to a close-packed dimer configuration (right), as explained in the text. The coupling \( \lambda \) is the fugacity of the empty sites in the worldline viewpoint or, equivalently, the fugacity of the inter-layer dimers in the dimer viewpoint.}
  \label{fig4}
\end{figure}

The challenge for qubit regularization is to reproduce the SSF using a lattice model with a finite-dimensional Hilbert space by appropriately tuning it to a critical point.  
While several such models are known to exist, an important feature of the qubit-regularized model studied in Ref.~\cite{AFPRL2024} is that the BKT critical point can be reached from the massive phase without the need for fine-tuning. In contrast to \cref{sec2}, where qubit regularization was implemented within the Hamiltonian formalism, Ref.~\cite{AFPRL2024} achieves it in the Euclidean space-time Lagrangian formulation by replacing the continuous degree of freedom \( \theta \) with a discrete degree of freedom that takes only a finite set of values. This Lagrangian formulation can also be understood through a transfer-matrix approach within a finite-dimensional Hilbert space.

\begin{figure}[t]
\centering \includegraphics[width=\textwidth]{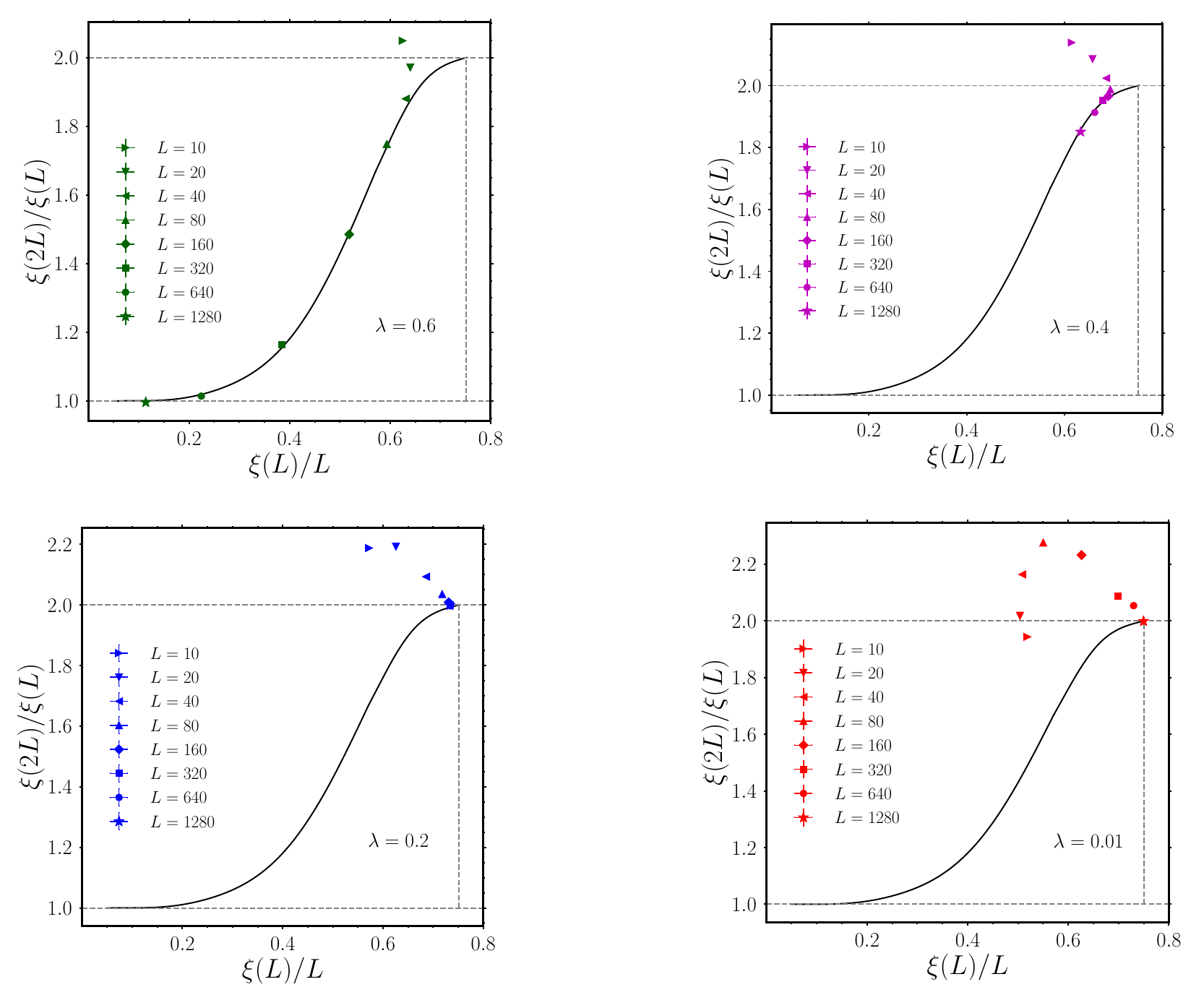}
\caption{Step-scaling function (SSF) of the 2D \( O(2) \) QFT, defined by tuning the traditional XY model to the BKT critical point from the massive phase. The solid line corresponds to the SSF obtained from the traditional XY model, while the data shown are from the qubit-regularized model described by \cref{dimer}. For $\lambda = 0.4, 0.6$, the data for $L > L_{\rm min}$ begins to follow the universal SSF of the 2D \( O(2) \) QFT. For $\lambda = 0.2, 0.01$, $L_{\rm min}$ is larger than the lattice sizes explored. These figures were originally published as supplementary material in \cite{AFPRL2024}.}
  \label{fig5}
\end{figure}

The Hilbert space of the traditional model,  
\( \cHT \), is a direct sum of irreps of the symmetry group, which in this case is the \( O(2) \) group. It can be expressed as  
\begin{align}
\cHT = \bigoplus_{m=0,\pm1,\pm2,\dots} \cH_m,    
\end{align}
where \( \cH_m \) corresponds to the angular momentum irreps.  

In contrast, the qubit regularization introduced in Ref.~\cite{AFPRL2024} uses a four-dimensional local Hilbert space given by  
\begin{align}
\cH_Q \ =\  \cH_{m=0} \ \oplus\  \cH_{m=0} \ \oplus\  \cH_{m=1}\ \oplus\ \cH_{m=-1}.    
\end{align}
Unlike \( \cHT \), the qubit-regularized Hilbert space includes two \( m=0 \) states, in addition to the states with angular momentum \( m=+1 \) and \( m=-1 \). As we will see, one of the \( m=0 \) states can be interpreted as a Fock vacuum state, while the other represents a bound state of \( m=+1 \) and \( m=-1 \). Consequently, this bound state still has a total angular momentum of zero.

The lattice model in this finite-dimensional Hilbert space can be described by the partition function  
\begin{align}
Z = \sum_{C} \lambda^{N_I},
\label{dimer}
\end{align}
where \( C \) represents worldline configurations of hardcore bosons carrying angular momentum \( +1 \) or \( -1 \) on a square lattice. An illustration of a configuration \( C \) is shown on the left side of \cref{fig4}. These configurations naturally take the form of closed, oriented loops that do not touch. In addition to sites containing particles, some sites can be empty, representing the absence of any particle and thus interpreted as Fock vacuum states. The quantity \( N_I \) in \cref{dimer} denotes the number of these empty sites in the configuration \( C \).  

With some effort, we can construct a transfer matrix for the partition function in \cref{dimer}, which has a four-dimensional local Hilbert space. To understand this, note that each lattice site is either empty or occupied by a particle with angular momentum \( m=+1 \) or \( m=-1 \), forming three distinct orthonormal states in the transfer matrix formulation. However, an additional state is required to describe loops that form across a single bond connecting neighboring sites. Each such pair of sites is interpreted as containing a bound state of two bosons with opposite angular momentum. These \( m=0 \) states constitute the fourth orthonormal basis state of the Hilbert space.  

As explained in Ref.~\cite{AFPRL2024}, the partition function in \cref{dimer} can also be viewed as a fermionic version of the XY model and is therefore referred to as the \( fXY \) model, in contrast to the traditional model, which is called the \( bXY \) model.

Interestingly, every worldline configuration \( C \) can be uniquely mapped to a configuration of close-packed oriented dimers on two layers of square lattices. To illustrate this mapping, the worldline configuration shown on the left of \cref{fig4} is mapped to the dimer configuration shown on the right side of \cref{fig4}. Note that the dimers are always oriented from even sites to odd sites. Since nearest-neighbor sites always have opposite parity, the site on the bottom layer has the opposite parity compared to the corresponding site on the top layer. Empty sites are mapped to inter-layer dimers that connect the two layers. 

In Ref.~\cite{AFPRL2024}, correlation functions of creation and annihilation operators for particles with angular momentum \( m=+1 \) and \( m=-1 \) were used to compute \( \xi(L) \) as a function of \( L \), which was then used to determine the SSF. The critical point of the qubit-regularized model that reproduces the SSF of the 2D \( O(2) \) QFT is found at \( \lambda=0 \). However, the RG flow through which this is achieved is once again described by \cref{fig2}.  

At \( \lambda=0 \), there are no inter-layer dimers, and the partition function given in \cref{dimer} describes the statistical mechanics of two decoupled layers of close-packed dimers. Such a system is known to be critical, describing free massless bosons. For large values of \( L \), one obtains \( \xi(L)/L \approx 0.4889(6) \), as seen in \cref{fig6}. This differs from the expected value at the BKT transition, where \( \xi(L)/L \approx  0.7506912... \), as explained above. Thus, the RG flow of the qubit-regularized model leads to a decoupled fixed point at \( \lambda=0 \) rather than the desired BKT UV fixed point.  

On the other hand, notice in \cref{fig6} that when \( \lambda=0.01 \), we observe \( \xi(L)/L \approx 0.7506912... \) on lattice sizes of \( L \approx 1000 \). This result can be understood by examining the SSF at different values of \( \lambda \), which are plotted in \cref{fig5}.  When \( \lambda=0.6 \) and \( 0.4 \), the data behaves similarly to what we discussed in \cref{sec2}. For \( L < L_{\rm min} \), the data does not follow the SSF of the 2D \( O(2) \) QFT, but for \( L > L_{\rm min} \), it begins to align with the desired curve.  

On the other hand, when \( \lambda=0.2 \) and \( 0.01 \), it appears that \( L_{\rm min} \) is larger than the largest lattice sizes we have studied. However, in these cases, \( \xi(L)/L \) approaches the BKT value of \( 0.7506912... \). This approach is most likely a crossover phenomenon, and for sufficiently large values of \( L \), the data will eventually begin to follow the expected SSF of the 2D \( O(2) \) QFT. 

All of this is once again consistent with the RG flow described by \cref{fig2}, which shows that asymptotically free UV fixed points can be recovered as a crossover criticality. Furthermore, the qubit-regularized model was able to recover the physics of the BKT critical point more easily than the traditional model, without the need for fine-tuning.

\section{Qubit Regularization of Gauge Theories}

In the two examples of qubit regularization discussed in the previous sections, asymptotically free QFTs emerged via a four-dimensional local Hilbert space through a novel RG flow that did not require fine-tuning. Can asymptotically free QFTs in higher dimensions, particularly non-Abelian gauge theories in 3+1 dimensions, also be formulated via qubit regularization with a small, finite-dimensional local Hilbert space? If this is possible, will these theories emerge through standard RG flows, as depicted in \cref{fig1}, or will they emerge via exotic flows, such as the one illustrated in \cref{fig2}? Answers to these questions have the potential to provide deeper insights into non-Abelian gauge theories, extending beyond the original motivation of studying these theories in the context of quantum computation.

\begin{figure*}[t]
\begin{center}
\hbox{
\includegraphics[width=0.4\textwidth]{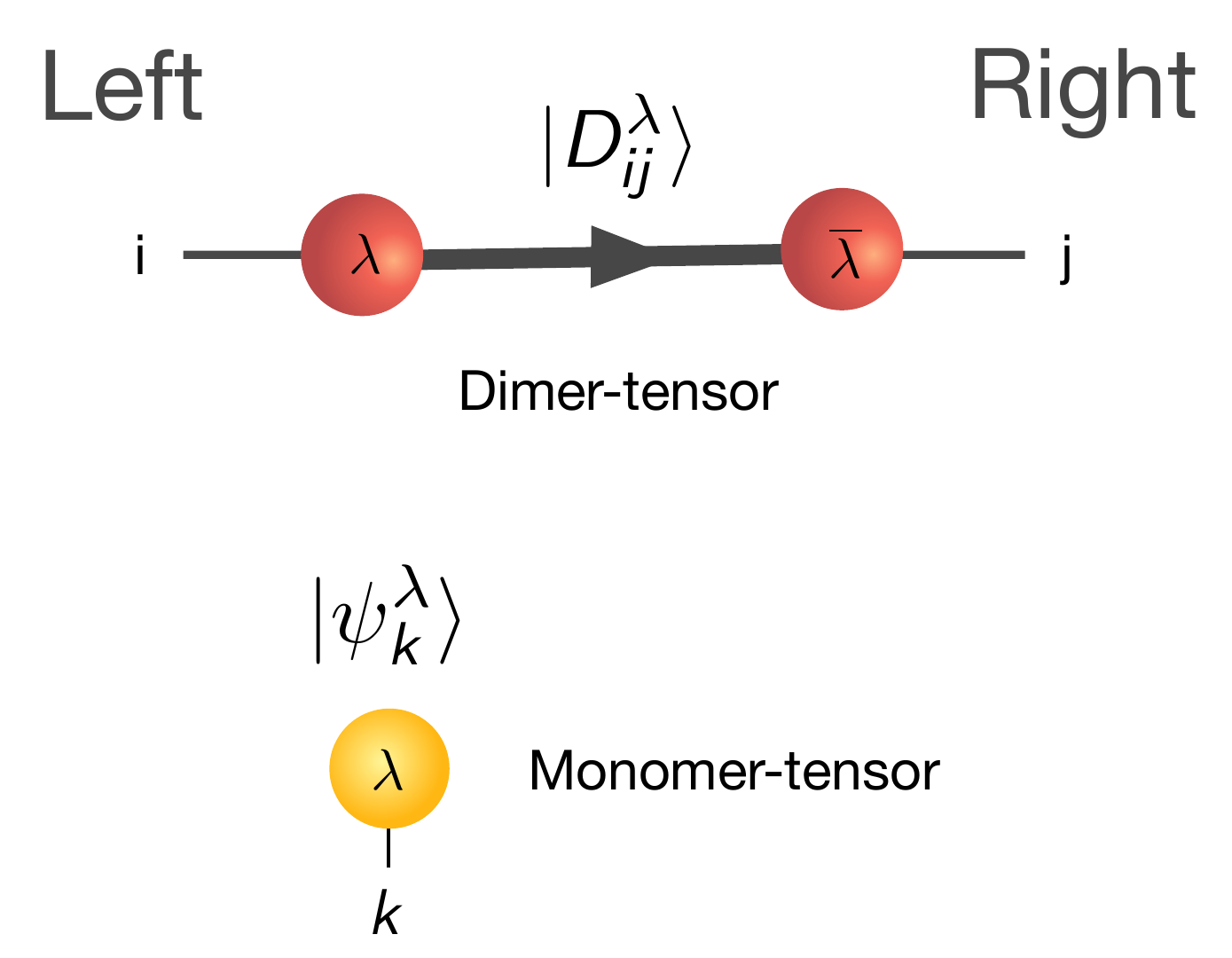} \hskip0.5in
\includegraphics[width=0.4\textwidth]{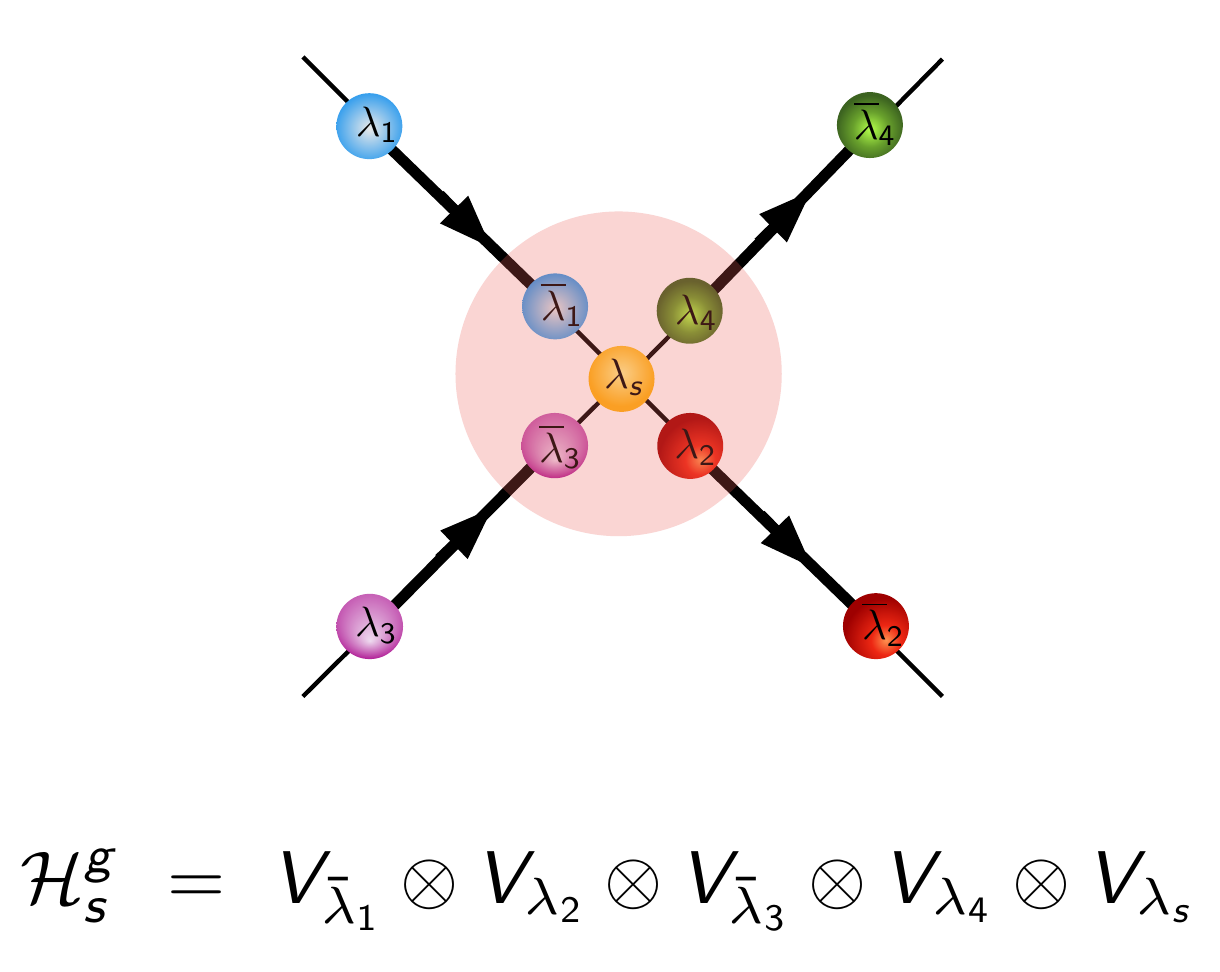}
}
\end{center} 
\caption{A dimer-tensor is associated with an oriented link and is represented by the irrep label $\lambda$ and the indices $i,j$, as shown at the top of the left figure. It represents the basis states $\ket{D^\lambda_{ij}}$. Under \(SU(N)\) gauge transformations, the index $i$ transforms as $V_\lambda$ and is associated with the left lattice site, while the index $j$ transforms as $V_{\bar{\lambda}}$ and is associated with the right lattice site. The monomer-tensor is associated with a site and is represented by the irrep label $\lambda$ as shown at the bottom of the left figure. It represents the basis states $\ket{\psi^\lambda_k}$, with index $k$ transforming as $V_\lambda$ under \(SU(N)\) gauge transformations. On each lattice site, the \(SU(N)\) gauge transformations act on the Hilbert space denoted as $\cH^g_s$, constructed with irreps $V_\lambda$'s from the links and the site associated with the site. An example is shown in the figure on the right. The index $\alpha_s = 1, 2, \dots, {\cal D}^g_s$ labels the singlet irreps of $\cH^g_s$.}
\label{fig7}
\end{figure*}

In this section, we build upon the discussion in \cref{sec2,sec3} by constructing new qubit-regularized gauge theories. While the foundations for our work were established years ago within the D-theory approach \cite{QLM1997,QLQCD1999}, several new ideas for qubit regularization continue to emerge (see, for example, \cite{LQCD2023,FuzGT2024} and references therein). Very little is known about quantum critical points and RG flows in qubit-regularized gauge theories. If, as commonly expected, the number of RG fixed points in higher dimensions is small, then recovering traditional gauge theories through qubit regularization may be relatively straightforward. However, the possibility of undiscovered exotic RG fixed points cannot be ruled out — an intriguing prospect in its own right. In any case, the first step in this pursuit is to construct qubit-regularized gauge theories and identify quantum critical points within them.

Qubit regularization of lattice gauge theories is providing new insights into the formulation of gauge theories themselves, encouraging us to view the physical Hilbert space of the theory through the lens of irreps of the gauge symmetry (see Ref.~\cite{Liu2022}). Building on this perspective, in this section, we will first construct an orthonormal basis for the physical Hilbert space of traditional lattice gauge theories that is well-suited for qubit regularization. We will develop a pictorial representation of these basis states and argue that they can be interpreted as a monomer-dimer-tensor-network (MDTN). We will then use the MDTN basis state perspective to guide us in constructing new types of qubit-regularized lattice gauge theories that can capture the physics of confinement and deconfinement.

Traditional lattice gauge theories contain quantum gauge degrees of freedom on the oriented links of the lattice and quantum matter degrees of freedom on the sites. The traditional link Hilbert space, \( \cH^{\rm Trad}_\ell \), is that of a quantum particle moving on the surface of the \( SU(N) \) manifold. As explained in Ref.~\cite{Liu2022}, if \( \lambda \) labels an irrep of \( SU(N) \), and \( V_\lambda \) denotes the corresponding Hilbert space with dimension \( d_\lambda \), we can use the Peter-Weyl theorem to write  
\begin{align}
    \cH^{\rm Trad}_\ell = \bigoplus_{\lambda_\ell} V_{\lambda_\ell} \otimes V_{\bar{\lambda}_\ell},
    \label{PWT}
\end{align}
where \(V_{\lambda_\ell} \otimes V_{\bar{\lambda}_\ell} \) is a \( d_{\lambda_\ell}^2 \)-dimensional subspace spanned by orthonormal basis states \( \ket{D^{\lambda_\ell}_{ij}} \), where \( i,j = 1,2,\dots,d_{\lambda_\ell} \), and each \( SU(N) \) irrep \( \lambda_\ell \) appears exactly once in the direct sum. Here, \( \bar{\lambda}_\ell \) denotes the conjugate representation of \( \lambda_\ell \). In the orthonormal basis states \( \ket{D^{\lambda_\ell}_{ij}} \) associated with the oriented link, the index \( i \) corresponds to degrees of freedom transforming under \( V_{\lambda_\ell} \) associated with the left lattice site, while \( j \) corresponds to those transforming under \( V_{\bar{\lambda}_\ell} \) associated with the right lattice site. Collectively, the basis states \( \ket{D^{\lambda_\ell}_{ij}} \) can be interpreted as a {\em tensor} associated with an {\em oriented dimer} on the link. A pictorial representation of this dimer-tensor is illustrated in \cref{fig7} (left).

Similarly, the traditional Hilbert space of matter degrees of freedom, \( \cH^{\rm Trad}_s \), can be decomposed in terms of irreps \( V_{\lambda_s} \), spanned by the basis states \( \ket{\psi^{\lambda_s}_k} \) for \( k = 1,2,\dots,d_{\lambda_s} \). For a fixed \( \lambda_s \), these basis states can be also be viewed as a tensor associated with a monomer on the site. A pictorial representation of this monomer-tensor is also illustrated in \cref{fig7}.

\begin{figure*}[t]
\begin{center}
\includegraphics[width=0.6\textwidth]{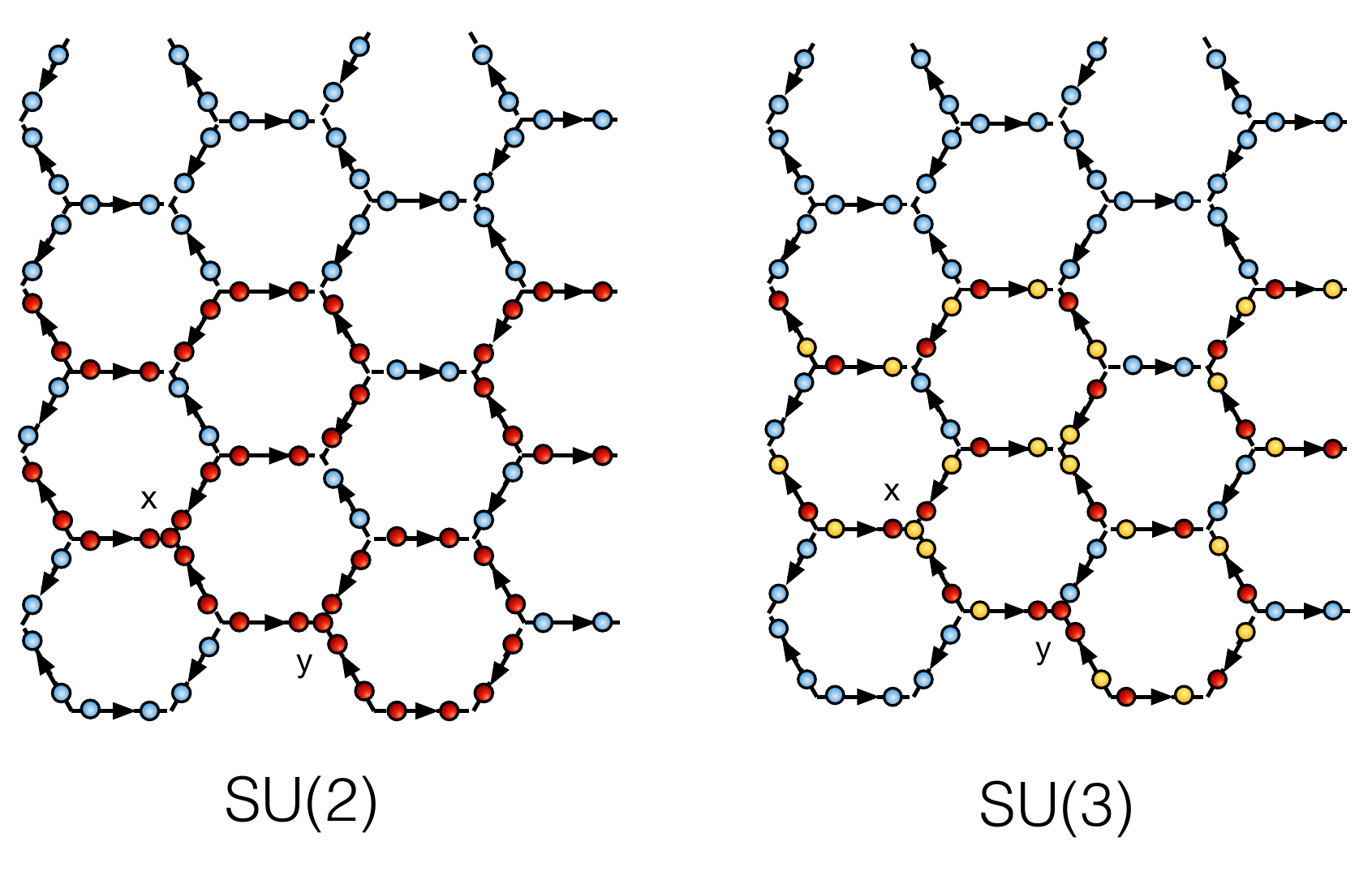}
\end{center} 
\caption{Pictorial representation of the MDTN basis states on honeycomb lattices in $SU(2)$ (left) and $SU(3)$ (right) lattice gauge theories in the ASQR scheme. The monomer-tensors with $\lambda_s=1$ have not been shown for convenience. Two sites $x$ and $y$ do contain matter fields with $\lambda_s\neq 1$. In the \(SU(2) \) case, $\lambda_s=2$ at both sites, while in the \(SU(3)\), we have $\lambda_y=3$ and $\lambda_x=\bar{3}$.}
\label{fig8}
\end{figure*}

For every fixed set of dimer-tensors on links \( \{\lambda_\ell\} \) and monomer-tensors on sites \( \{\lambda_s\} \), the physical Hilbert space of the lattice gauge theory is obtained by projecting onto the subspace of gauge-invariant states. This is commonly referred to as imposing Gauss's law. This projection can be achieved by constructing the Hilbert space \( \mathcal{H}_s^g \) at each lattice site, on which the gauge transformations act. This space is the direct product of all irreps \( V_{\lambda} \) associated with the site. An illustration of \( \mathcal{H}_s^g \) is shown in \cref{fig7}. The gauge-invariant singlet subspace is obtained by appropriate tensor contractions (or fusion rules) on the indices of the \( V_{\lambda} \)'s contained in \( \mathcal{H}_s^g \). If the dimension of this singlet space is denoted as \( \mathcal{D}(\mathcal{H}_s^g) \), we can use an index \( \alpha_s = 1, 2, \dots, \mathcal{D}(\mathcal{H}_s^g) \) to label the different orthonormal basis states of the physical Hilbert space on that site.

We observe that an orthonormal basis of the physical Hilbert space of a traditional \( SU(N) \) lattice gauge theory can be constructed using the set of monomer and dimer tensors labeled \( \{\lambda_\ell\} \) and \( \{\lambda_s\} \), along with the set \( \{\alpha_s\} \) that denotes the fusion rules used to construct the gauge-invariant states. We denote this orthonormal basis of a traditional lattice gauge theory as \( \ket{\mdc} \) and represent it pictorially as a monomer-dimer-tensor network (MDTN). For more details on the MDTN basis states, we refer the reader to Ref.~\cite{MDTN2025}, that should be published shortly.

Qubit-regularized lattice gauge theories can be constructed using the MDTN basis states by simply restricting the values of \( \lambda_\ell \) on the links in \( \ket{\mdc} \). While more complex qubit regularization schemes are possible, in Ref.~\cite{Liu2022}, a simple qubit regularization scheme was proposed by restricting \( \lambda_\ell \) to the anti-symmetric irreps of \( SU(N) \). We refer to this as the anti-symmetric qubit regularization (ASQR) scheme. In \cref{fig8}, we illustrate a monomer-dimer-tensor-network basis state for both \( SU(2) \) and \( SU(3) \) gauge theories in the ASQR scheme. We will argue in the next section that even the simple ASQR scheme is able to capture the finite-temperature confinement-deconfinement physics of traditional lattice gauge theories.

\section{Confinement-Deconfinement Transitions}

Using the MDTN basis states \( \ket{\mdc} \), we can explore lattice gauge theories from a new perspective. In particular, by employing them, we can construct new local Hamiltonians that are free of the sign problem, eliminating the need to resort to the original Kogut-Susskind approach \cite{KS1975}, which often suffers from sign problems due to the introduction of Clebsch-Gordan coefficients.

A key question is whether these non-traditional Hamiltonians can host quantum critical points where continuum non-Abelian gauge theories emerge through interesting RG flows. A small but significant first step toward this goal is to recover the finite-temperature physics of traditional $SU(N)$ lattice gauge theories, which exhibit a confined phase at low temperatures and a deconfined phase at high temperatures. In $d$ spatial dimensions, these classical transitions are known to follow the order-disorder physics of $Z_N$ spin models, where the low-temperature confined phase corresponds to the disordered phase, while the high-temperature deconfined phase corresponds to the ordered phase \cite{SY1982}.

\begin{figure*}[t]
\begin{center}
\includegraphics[width=\textwidth]{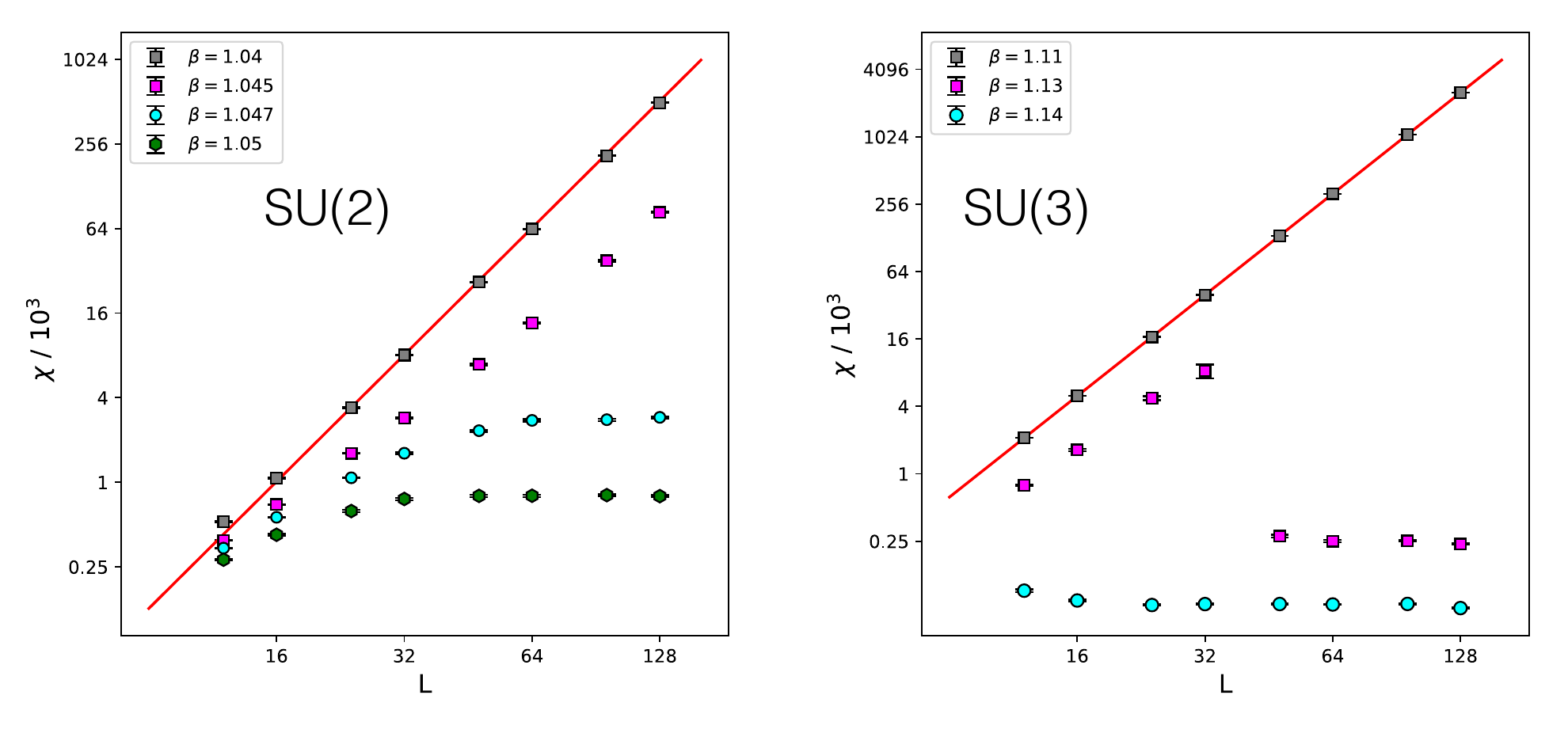}
\end{center} 
\caption{Plot of $\chi$ as a function of the lattice size $L$, showing the confinement-deconfinement transition as a function of $\beta$ on the diamond lattice ($d=3$) in qubit-regularized lattice gauge theory within the ASQR scheme. The left plot shows results for the $SU(2)$ gauge theory, which is consistent with the 3D Ising transition, while the right plot shows the transition in the $SU(3)$ gauge theory, which is first order.}
\label{fig9}
\end{figure*}

To recover classical transitions, it is usually sufficient to begin with a classical Hamiltonian in which every MDTN basis state is an eigenstate. A simple choice is $
H(\chE) = \sum_\ell \chE_\ell$, where  
$\chE_\ell \ket{\mdc} = (1-\delta_{\lambda,1}) \ket{\mdc}$.  
The energy associated with each MDTN basis state is then given by ${\cal E}(\mdc) = \sum_\ell (1-\delta_{\lambda_\ell,1})$.  

To study the finite-temperature confinement-deconfinement phase transition using this classical Hamiltonian in the ASQR scheme, we focus on the pure gauge theory with \(\lambda_s = 1\) on all lattice sites. Additionally, we also consider a system with two sites, \(x\) and \(y\), where heavy matter fields are introduced so that \(\lambda_x\) and \(\lambda_y\) belong to either the fundamental or anti-fundamental irrep of \(SU(N)\). Let \(Z\) and \(Z^{(x,y)}\) denote the corresponding partition functions in the pure gauge sector and with heavy matter fields at \(x\) and \(y\). Using these, we define the susceptibility as  
\begin{align}
\chi = \frac{1}{L^d} \sum_{x,y} \frac{Z^{(x,y)}}{Z},     
\end{align}  
where \(L^d\) is the spatial lattice volume. In the thermodynamic limit, \(\chi\) is expected to approach a constant in the confined phase and scale with \(L^d\) in the deconfined phase.  

We computed \(\chi\) as a function of lattice size \(L\) for \(d=2\) (honeycomb lattice) and \(d=3\) (diamond lattice) using loop Monte Carlo algorithms \cite{ALGO2003}. Our results for the phase transitions align with expectations from traditional lattice gauge theories. For instance, in \(SU(2)\) gauge theories, the confined-to-deconfined phase transition is second order and belongs to the Ising universality class. In \(SU(3)\) gauge theories, the transition follows the three-state Potts model in \(d=2\), while in \(d=3\), it is first order. In \cref{fig9}, we present our results for the \(d=3\) case in \(SU(2)\) (left) and \(SU(3)\) (right) gauge theories. Further details can be found in \cite{MDTN2025}.  

\section{Conclusions}

Qubit regularization of quantum field theories, while initially motivated by quantum computing, offers new opportunities to gain deeper insights beyond the realm of quantum computing. We have already uncovered new RG flows that can exactly recover asymptotic freedom in qubit-regularized models with a finite local Hilbert space. Additionally, we have learned how to formulate lattice gauge theories using the irreps of the gauge symmetry. These orthonormal basis states, which we refer to as MDTN basis states, encourage us to rethink gauge theories from a fresh perspective that extends beyond the perturbative approach introduced long ago. 

The new Hamiltonians can be constructed to be sign-problem-free, unlike the traditional approaches, which suffer from these issues related to Clebsch-Gordan coefficients. Recent work in one dimension already shows hints that the string tension can slowly be tuned to zero within these new quantum Hamiltonians \cite{MDTN2025}. The exploration of quantum critical points and RG flows in these new qubit-regularized gauge theories in higher dimensions promises to be a fruitful and compelling research direction for the future.

\section*{Acknowledgments}

This work brings together ideas developed through multiple collaborations over the years. I thank all my collaborators, particularly T. Bhattacharya, H. Liu, R. X. Siew, and U.-J. Wiese, for sharing their valuable insights with me. We acknowledge the use of AI assistance, specifically ChatGPT, in refining the language and clarity of this manuscript. This research was supported in part by a Duke subcontract of the U.S. Department of Energy, Office of Science, High Energy Physics Contract KA2401032 to Los Alamos National Laboratory, and by the U.S. Department of Energy, Office of Science, Nuclear Physics program under Award No. DE-FG02-05ER41368.

\end{document}